# Characteristic function and quasi-probability distribution of photons and of squeezed coherent photons

Moorad Alexanian

*Department of Physics and Physical Oceanography*
*University of North Carolina Wilmington*
*Wilmington, NC 28403-5606*

**Abstract:** We consider the *p*-ordered characteristic function and its Fourier transform, the quasi-distribution function, of squeezed coherent photons in a thermal state of photons and calculate the mean number and number variance of squeezed coherent photons. All the properties calculated from the previous characteristic function can be used to calculate said properties for photons in a thermal state of squeezed coherent photons via a parametric transformation. In particular, one can obtain the mean number and number variance via the parametric transformation.

## 1. Introduction

There are three basic states in quantum optics and in a recent paper [1], we dealt with the quantum Rabi oscillations via one- and two-photon transitions in two of these basic states, viz., coherent and squeezed coherent states. More recently we dealt with the third state, viz., thermal states, and studied the mean number and number variance of squeezed coherent photons in a thermal state of photons [2]. In the present paper, we calculate the characteristic function and the quasi-probability distribution of photons in a thermal state of squeezed coherent photons and, conversely, the case of squeezed coherent photons in a thermal state of photons. As expected, the latter is related to the former via a parametric transformation.

This paper is arranged as follows. In Sec. 2, we introduce the *p*-ordered characteristic function and determine it for squeezed coherent photons in a thermal state of photons. In Sec. 3, we calculate the quasi-probability function, which is the Fourier transform of the characteristic function. In Sec. 4, we calculate the mean number and number variance of squeezed coherent photons in a thermal state of photons. In Sec. 5, we repeat the previous calculations but for photons in a thermal state of squeezed coherent photons. Finally, Sec. 5 summarizes our results.

## 2. Characteristic function

The creation and annihilation operators $\hat{B}$ and $\hat{B}^{\dagger}$, respectively, for the squeezed coherent photons are given by

$$\hat{B} = \hat{S}(\zeta)\hat{D}(\alpha)\hat{a}\hat{D}(-\alpha)\hat{S}(-\zeta) = \cosh(r)\hat{a} + e^{i\varphi}\sinh(r) - \alpha \qquad (1)$$

and

$$\hat{B}^\dagger = \hat{S}(\zeta)\hat{D}(\alpha)\hat{a}^\dagger \hat{D}(-\alpha)\hat{S}(-\zeta) = e^{-i\varphi}\sinh(r)\hat{a} + \cosh(r)\hat{a}^\dagger - \alpha^*, \qquad (2)$$

where the Glauber displacement operator is given by

$$\hat{D}(\alpha) = \exp(\alpha \hat{a}^\dagger - \alpha^* \hat{a}) \qquad (3)$$

and the squeezing operator by

$$\hat{S}(\zeta) = \exp\left(-\frac{\zeta}{2}\hat{a}^{\dagger 2} + \frac{\zeta^*}{2}\hat{a}^2\right) \qquad (4)$$

with $\zeta = r\exp(i\varphi)$.

The *p*-ordered characteristic function of squeezed coherent photons in a thermal state of photons is

$$\chi_a(\xi, p) = Tr\left[\hat{\rho}_a \exp\left(\xi \hat{B}^\dagger - \xi^* \hat{B}\right)\right]\exp\left(p|\xi|^2/2\right), \qquad (5)$$

where

$$\hat{\rho}_a = \frac{\exp\left(-\beta_a \hbar \omega \hat{a}^\dagger \hat{a}\right)}{Tr\left[\exp\left(-\beta_a \hbar \omega \hat{a}^\dagger \hat{a}\right)\right]}, \qquad (6)$$

with $\beta_B = (k_B T)^{-1}$, $k_B$ being Boltzmann's constant and $T$ being the absolute temperature.

One obtains

$$\chi_a(\xi, p) = \exp\left(-(\bar{n}+1/2)\left|\xi\cosh(r) - \xi^*\sinh(r)e^{i\varphi}\right|^2 + p|\xi|^2/2 - \xi\alpha^* + \xi^*\alpha\right), \qquad (7)$$

where

$$\langle \hat{a}^\dagger \hat{a}\rangle = \bar{n} = \left[\exp(\beta \hbar \omega) - 1\right]^{-1} \qquad (8)$$

and use has been made of the characteristic function of a thermal state [3] of photons with mean number $\bar{n}$ which is

$$\chi(\xi, p) = \exp\left[-(\bar{n}+1/2 - p/2)|\xi|^2\right]. \qquad (9)$$

Note that the characteristic function (7) depends on both the modulus and argument of $\xi$ this is so since the density matrix in (5), viz., $\hat{D}(-\alpha)\hat{S}(-\zeta)\hat{\rho}_a\hat{S}(\zeta)\hat{D}(\alpha)$, is not diagonal in the number state basis.

The expectation value of the *p*-ordered product of $\hat{B}^{\dagger m}$ and $\hat{B}^n$ is

$$\left\langle \hat{B}^{\dagger m}\hat{B}^n \right\rangle_p = \left(\frac{\partial}{\partial \xi}\right)^m \left(-\frac{\partial}{\partial \xi^*}\right)^n \chi_a(\xi, p)\Bigg|_{\xi=0}. \tag{10}$$

This equation also serves as the definition of *p*-ordering. For $p=1$ and $p=-1$, we obtain $\left\langle \hat{B}^{\dagger m}\hat{B}^n \right\rangle$ and $\left\langle \hat{B}^m \hat{B}^{\dagger n} \right\rangle$, respectively, while for $p=0$ we obtain the expectation value of the symmetrized product $\left\langle S\left(\hat{B}^{\dagger m}\hat{B}^n\right)\right\rangle$ [3].

### 3. Quasi-probability distribution

The quasi-probability distribution $W(\eta, p)$ corresponding to the *p*-ordered characteristic function $\chi_a(\xi, p)$ is

$$W_a(\eta, p) = \frac{1}{\pi^2}\int d^2\xi \chi_a(\xi, p)\exp(\eta \xi^* - \eta^*\xi) \tag{11}$$

which becomes, with the aid of (1), (2) and (7),

$$W_a(\eta, p) = \frac{2}{\pi\sqrt{4A^2B^2-C^2}}\exp\left(-\frac{A^2F(e1)^2 + B^2E(e2)^2 + CE(e2)F(e1)}{4A^2B^2-C^2}\right), \tag{12}$$

where

$$A^2 = -(\bar{n}+1/2)\cos(\varphi)\sinh(2r) + \bar{n}\cosh(2r) + \sinh^2(r) + \frac{1}{2}(1-p) \tag{13}$$

$$B^2 = (\bar{n}+1/2)\cos(\varphi)\sinh(2r) + \bar{n}\cosh(2r) + \sinh^2(r) + \frac{1}{2}(1-p), \tag{14}$$

$$C = (2\bar{n}+1)\sin(\varphi)\sinh(2r), \tag{15}$$

$$E(e2) = 2a2 + 2e2 \qquad F(e1) = -2a1 - 2e1, \tag{16}$$

and

$$\alpha = a1 + ia2 \qquad \eta = e1 + ie2. \tag{17}$$

In the two dimensional integral (11), $\xi = x + iy$, $d^2\xi = dxdy$, and $-\infty \leq x, y \leq \infty$.

The usual values p=1,0,1 correspond to quasi-probability distributions giving normal, symmetric, or anti-normal ordered moments, respectively. The Glauber-Sudarshan *P*-representation corresponds to $W(\eta,1) = P(\eta)$. The Wigner function is obtained for $p = 0$, viz., $W(\eta,0) = W(\eta)$. The Husimi or $Q-$function is given by $W(\eta,-1) = Q(\eta)$ [3]. Unlike the $P-$function, the $Q-$function has all the properties of a classical probability distribution whereas the $P-$function can be singular and need not exist. The Wigner function $W(\eta)$ does not have the singular behavior sometimes found for $P(\eta)$, but it can assume negative values [3].

### 4. Mean number and number variance

The quasi-probability distribution $W_a(\eta, p)$ is normalized to unity

$$\int d^2\eta W_a(\eta, p) = 1 \qquad (18)$$

and moments of the desired *p*-ordered product of $\hat{B}$ and $\hat{B}^\dagger$ can be obtained by evaluating the integral

$$\left\langle \hat{B}^{\dagger m} \hat{B}^n \right\rangle_p = \int d^2\eta W_a(\eta, p) \eta^{*m} \eta^n. \qquad (19)$$

We obtain, with the aid of (12), for the mean number of squeezed coherent photons that

$$\left\langle \hat{B}^\dagger \hat{B} \right\rangle_p = \int d^2\eta W_a(\eta,p)|\eta|^2 = \frac{\cosh(2r)}{e^{\beta_B \hbar \omega} - 1} + \sinh^2(r) + |\alpha|^2 + \frac{1}{2}(1-p). \qquad (20)$$

Similarly, the number variance is given by

$$\left\langle \hat{B}^{\dagger 2}\hat{B}^2 \right\rangle_p + \left\langle \hat{B}^\dagger \hat{B} \right\rangle_p - \left\langle \hat{B}^\dagger \hat{B} \right\rangle_p^2 = \frac{\cosh(4r)}{(e^{\beta_B \hbar \omega}-1)^2} + \frac{\cosh(4r) + 2\left|\alpha\cosh(r)+\alpha^*\sinh(r)e^{i\varphi}\right|^2}{(e^{\beta_B \hbar \omega}-1)} +$$
$$+ \frac{1}{2}\sinh^2(2r) + \left|\alpha\cosh(r)+\alpha^*\sinh(r)e^{i\varphi}\right|^2 + \frac{1}{4}(1-p)^2 + (1-p)\left[(n+1/2)\cosh(2r)+|\alpha|^2\right]. \qquad (21)$$

The results for the mean number and number variance in Ref.[2] follow from (20) and (21) for $p = 1$ when

$$\Delta n^2 = \left\langle \left(\hat{B}^\dagger \hat{B}\right)^2 \right\rangle - \left\langle \hat{B}^\dagger \hat{B} \right\rangle^2. \qquad (22)$$

## 5. Thermal state of squeezed coherent photons

The *p*-ordered characteristic function of photons in a thermal state of squeezed coherent photons is

$$\chi_B(\xi, p) = Tr\left[\hat{\rho}_B \exp(\xi \hat{a}^\dagger - \xi^* \hat{a})\right] \exp\left(p|\xi|^2/2\right), \tag{23}$$

where

$$\hat{\rho}_B = \frac{\exp(-\beta_B \hbar \omega \hat{B}^\dagger \hat{B})}{Tr\left[\exp(-\beta_B \hbar \omega \hat{B}^\dagger \hat{B})\right]}. \tag{24}$$

One obtain for the characteristic function

$$\chi_B(\xi, p) = \exp\left(-(\bar{n}+1/2)|\xi\cosh(r) + \xi^*\sinh(r)e^{i\varphi}|^2 + p|\xi|^2/2\right) \times$$
$$\times \exp\left(\xi(\alpha^*\cosh(r) - \alpha\sinh(r)e^{-i\varphi}) - \xi^*(\alpha\cosh(r) - \alpha^*\sinh(r)e^{i\varphi})\right), \tag{25}$$

with the aid of (1), (2) and (7). Note that the characteristic function $\chi_B(\xi, p)$ follows from $\chi_a(\xi, p)$ under the transformation

$$\alpha \to -(\alpha \cosh(r) - \alpha^* \sinh(r) e^{i\varphi}) \quad \text{and} \quad \zeta \to -\zeta. \tag{26}$$

The parametric transformation (26) is generated from the density matrix $\hat{\rho}_a$ by the unitary transformation $\hat{\rho}_B = \hat{U}\hat{\rho}_a \hat{U}^\dagger$, where $\hat{U} = \hat{S}(\zeta)\hat{D}(\alpha)$. Therefore, (20) and (21) yield under the transformation (26)

$$\langle \hat{a}^\dagger \hat{a} \rangle_p = \frac{\cosh(2r)}{e^{\beta_B \hbar \omega} - 1} + \sinh^2(r) + |\alpha \cosh(r) - \alpha^* \sinh(r) e^{i\varphi}|^2 + \frac{1}{2}(1-p) \tag{27}$$

and

$$\langle \hat{a}^{\dagger 2} \hat{a}^2 \rangle_p + \langle \hat{a}^\dagger \hat{a} \rangle_p - \langle \hat{a}^\dagger \hat{a} \rangle_p^2 = \frac{\cosh(4r)}{(e^{\beta_B \hbar \omega} - 1)^2} + \frac{\cosh(4r) + 2|\alpha \cosh(2r) - \alpha^* \sinh(2r)e^{i\varphi}|^2}{(e^{\beta_B \hbar \omega} - 1)} + \frac{1}{2}\sinh^2(2r) +$$
$$+ |\alpha \cosh(2r) - \alpha^* \sinh(2r)e^{i\varphi}|^2 + \frac{1}{4}(1-p)^2 + (1-p)\left[(n+1/2)\cosh(2r) + |\alpha \cosh(r) - \alpha^* \sinh(r)e^{i\varphi}|^2\right]. \tag{28}$$

The results for the mean number and number variance in Ref.[2] follow from (27) and (28) for $p = 1$ when

$$\Delta n^2 = \langle (\hat{a}^\dagger \hat{a})^2 \rangle - \langle \hat{a}^\dagger \hat{a} \rangle^2. \tag{29}$$

## 6. Summary and discussion

We have presented results for the *p*-ordered characteristic function and its Fourier transform, the quasi-probability distribution, of photons and of squeezed photons. We have calculated the mean number and number variance of photons in a thermal state of squeezed photons and, conversely, of squeezed coherent photons in a thermal state of photons. One can obtain the latter from the former via a parametric transformation.